%% file: proc4.tex
\documentclass[12pt]{article}
\usepackage{a4wide}
\usepackage{graphicx}
\def\1{\mbox{l\hspace{-0.53em}1}}
\DeclareMathAlphabet{\mathbbm}{U}{bbm}{m}{n}
  \SetMathAlphabet\mathbbm{bold}{U}{bbm}{bx}{n}

\begin{document}
\def\d{{\rm d}}
\title{\bf Highly excited states of baryons in large $N_c$ QCD}
\author{N. Matagne
\thanks{{\it e-mail}: nicolas.matagne@umons.ac.be} ~and Fl.\ Stancu\thanks{{\it e-mail}: fstancu@ulg.ac.be} \\
 {\small $^\ast$ Service de Physique Nucl\'eaire et Subnucl\'eaire, University of Mons,}\\[-6pt]
 {\small Place du Parc, B-7000 Mons, Belgium,} \\ %
 {\small $^\dagger$ Institute of Physics, B5, University of Li\`ege,}\\[-6pt]
 {\small Sart Tilman, B-4000 Li\`ege 1, Belgium}}
\maketitle
%%%%%%%%%%%%%
\begin{abstract}\noindent
The masses of highly excited negative parity baryons belonging to the $N$ = 3 band are calculated in the 
$1/N_c$ expansion method of QCD. We use a procedure which allows to write the mass formula by 
using a small number of linearly independent operators. The numerical fit of the dynamical coefficients
in the mass formula show that the pure spin and pure flavor terms are dominant in the expansion, like for the $N$ = 1 band.
We present the trend of some important dynamical coefficients as a function of the band number $N$ or 
alternatively of the excitation energy.

\end{abstract}

%%%%%
\section{The status of the $1/N_c$ expansion method}

The large $N_c$ QCD, or alternatively the $1/N_c$ expansion method, proposed by  't Hooft  \cite{HOOFT} in 1974
and implemented  by Witten  in 1979 \cite{WITTEN} became  a valuable tool to study baryon properties 
in terms of the parameter $1/N_c$  where $N_c$ is the number
of colors.  
According to Witten's intuitive picture, a baryon containing $N_c$ quarks 
is seen as a bound state in an average self-consistent potential of a Hartree type 
and the corrections to the Hartree approximation are of order $1/N_c$.
These corrections capture the key phenomenological features of the baryon structure.

Ten years after 't Hooft's work, Gervais and Sakita  \cite{Gervais:1983wq}
and independently Dashen and Manohar in 1993  \cite{DM} derived a set of consistency conditions for the pion-baryon
coupling constants  which imply that  the large $N_c$ limit of QCD 
has an exact contracted SU(2$N_f$)$_c$ symmetry  
when $N_c \rightarrow \infty $,   $N_f$ being the number 
of flavors.
For ground state baryons the SU(2$N_f$) symmetry is broken by 
corrections proportional to $1/N_c$
\cite{Dashen:1994qi,Jenkins:1998wy}.

Analogous to s-wave baryons, consistency conditions which constrain the strong couplings  
of excited baryons to pions were derived  in Ref. \cite{Pirjol:1997bp}.
These consistency conditions predict the equality between  pion couplings to excited states
and pion couplings to s-wave baryons. These predictions are consistent with the nonrelativistic 
quark model.

A few years later, in the spirit of the Hartree approximation
a procedure for constructing  large $N_c$ baryon wave functions 
with mixed symmetric spin-flavor parts has been proposed
\cite{Goity:1996hk} and an operator analysis was performed for $\ell$ = 1
baryons  \cite{Carlson:1998vx}. 
It was proven that,  for such states,
the SU($2N_f$) breaking occurs at order $N^0_c$, instead of $1/N_c$, as it is the case for  ground and also for symmetric 
excited states  $[56, \ell^+]$ (for the latter see Refs. \cite{Goity:2003ab,Matagne:2004pm}).
This procedure has been extended to positive parity nonstrange baryons belonging to the $[70, \ell^+]$ multiplets with $\ell$ = 0 and 2
\cite{Matagne:2005gd}. In addition, in Ref.  \cite{Matagne:2005gd},  the dependence of the contribution of the linear term in $N_c$, of the spin-orbit
and of the spin-spin terms in the mass formula was  presented as a function of the excitation energy
or alternatively in terms of the band number $N$.  
Based on this analysis   an impressive global compatibility between the $1/N_c$ expansion and the quark model results  
for $N$ = 0, 1, 2 and 4  was found \cite{Semay:2007cv}
(for a review see Ref. \cite{Buisseret:2008tq}).
More recently the $[70,1^-]$ multiplet was reanalyzed by using an exact wave function, instead of the 
Hartree-type wave function, which allowed to keep control of the Pauli principle at any stage 
of the calculations \cite{Matagne:2006dj}.  The novelty was that the isospin term, neglected previously
 \cite{Carlson:1998vx} becomes as dominant in $\Delta$ resonances as the spin term in $N^*$ resonances.

The purpose of this work is mainly to complete the analysis of the excited states by including the $N = 3$ band for which results 
were missing in the systematic analysis of Ref. \cite{Matagne:2005gd}. An incentive for studying highly excited states 
with $\ell$ = 3 has been given by a recent paper \cite{Matagne:2011sn}
where the compatibility between the two alternative pictures for baryon resonances namely the ${\it quark-shell ~picture}$  
and the ${\it meson-nucleon ~scattering}$ picture defined in the framework of chiral soliton models
\cite{HAYASHI,MAPE} has been proven explicitly. This work was an extension of the 
analysis made independently  by Cohen and Lebed \cite{COLEB1,COLEB2} and Pirjol and Schat \cite{Pirjol:2003ye}
for low excited states with $\ell$ = 1. 

As explained below, we shall analyze the resonances thought to belong to the $N$ = 3 band  by using the procedure we have proposed in Ref.
\cite{Matagne:2006dj} for the $N$ = 1 band. Details can be found in Ref. \cite{Matagne:2012tm}.  

%%%%%%%%%%%%%%%%%%%%%%%%%%%%%%%%%%%%%%%%%%%%%%%%%%%%%%%%%%%
\section{Mixed symmetric baryon states}\label{se:excit}

If an excited baryon belongs to a symmetric SU(6) multiplet
the $N_c$-quark system  can be treated similarly to the ground state
in the flavour-spin degrees of freedom, but one has to take into
account the presence of an orbital excitation in the space
part of the wave function  \cite{Goity:2003ab,Matagne:2004pm}.
If the baryon state is described by 
a mixed symmetric representation of SU(6) , the $[\bf{70}]$ at $N_c$ = 3, 
the treatment becomes more complicated. 
In particular, the 
resonances up to about 2 GeV are thought to belong to  $[{\bf 70},1^-]$, $[{\bf 70},0^+]$ or 
$[{\bf 70},2^+]$ multiplets and beyond to 2 GeV to  $[{\bf 70},3^-]$, $[{\bf 70},5^-]$, etc.

There are two ways of studying mixed symmetric multiplets. 
The standard one is inspired by the Hartree approximation \cite{Goity:1996hk} where an
excited baryon is described by a symmetric core plus 
an excited quark coupled to this core, see \emph{e.g.} 
\cite{Carlson:1998vx,Matagne:2005gd,Goity:2002pu,Matagne:2006zf}.
The core is treated in a way similar to that of the ground state.
In this method each SU($2N_f$) $\times$ O(3) generator is  separated
into two parts 
\begin{equation}\label{CORE}
S^i = s^i + S^i_c; ~~~~T^a = t^a + T^a_c; ~~~ G^{ia} = g^{ia} + G^{ia}_c;
~~~ \ell^i = \ell^i_q + \ell^i_c,
\end{equation}
%and for the SO(3) generators $\ell^i$ one has 
%\begin{equation}\label{SO3}
%\ell^i = \ell^i_q + \ell^i_c,
%\end{equation}
where  $s^i$, $t^a$, $g^{ia}$ and $\ell^i_q$  are the excited 
quark operators and  
$S^i_c$, $T^a_c$, $G^{ia}_c$ and  $\ell^i_c$ the corresponding core operators.

As an alternative,   we have proposed a method where 
all identical quarks are treated on the same footing and we have an exact wave 
function in the orbital-flavor-spin space. The procedure has been successfully applied to the 
$N$ = 1 band  \cite{Matagne:2006dj,Matagne:2008kb,Matagne:2011fr}.
In the following we shall adopt this procedure to analyze the $N$ = 3 band.

%%%%%%%%%%%%%%%%%%%%%%%%%%%%%%%%%%%%%%%%%%%%%

\section{The mass operator}

When  hyperons are included in the analysis, the SU(3) symmetry must be broken and the mass operator takes the following general 
form \cite{JL95} 
\begin{equation}
\label{massoperator}
M = \sum_{i}c_i O_i + \sum_{i}d_i B_i .
\end{equation} 
The formula contains two types of operators. The first type are the operators $O_i$,  which are 
invariant under  SU($N_f$) and are defined as  % (\ref{OLFS})
\begin{equation}\label{OLFS}
O_i = \frac{1}{N^{n-1}_c} O^{(k)}_{\ell} \cdot O^{(k)}_{SF},
\end{equation}
where  $O^{(k)}_{\ell}$ is a $k$-rank tensor in SO(3) and  $O^{(k)}_{SF}$
a $k$-rank tensor in SU(2)-spin.  Thus $O_i$ are rotational invariant.
For the ground state one has $k = 0$. The excited
states also require  $k = 1$ and $k = 2$ terms. 
The rank $k = 2$ tensor operator of SO(3) is
\begin{equation}\label{TENSOR} 
L^{(2)ij} = \frac{1}{2}\left\{L^i,L^j\right\}-\frac{1}{3}
\delta_{i,-j}\vec{L}\cdot\vec{L},
\end{equation}
which we choose to act on the orbital wave function $|\ell m_{\ell} \rangle$  
of the whole system of $N_c$ quarks (see  Ref. \cite{Matagne:2005gd} for the normalization 
of $L^{(2)ij}$). 
The second type are the operators  $B_i$ which are 
SU(3) breaking and are defined to have zero expectation values for non-strange baryons.
Due to the scarcity of data in the $N$ = 3 band hyperons,  here we consider only one four-star hyperon
$\Lambda(2100)7/2^-$ and 
accordingly include only one of these operators,
namely $B_1 = - \mathcal{S}$ where $\mathcal{S}$ is the strangeness.

\begin{table*}[h!]%resp2_ed
\caption{Operators and their coefficients in the mass formula obtained from 
numerical fits. The values of $c_i$ and $d_i$ are indicated under the heading Fit $n\ (n=1,2,3,4)$ from Ref. \cite{Matagne:2012tm}.}
%In both fits  the resonance $N(2190)7/2^-$ is interpreted as a member of the $^2N(70,3^-)$ multiplet.}
\label{operators}{\scriptsize
\renewcommand{\arraystretch}{2} % enlarge line spacing
\begin{tabular}{lrrrr}
\hline
\hline
Operator \hspace{2cm} &\hspace{0.0cm} Fit 1 (MeV) & \hspace{1.cm} Fit 2 (MeV)  & \hspace{1.cm} Fit 3 (MeV)  & \hspace{1.cm} Fit 4 (MeV)  \\
\hline
%\hline%
$O_1 = N_c \ \1 $                                                & $c_1 = 672 \pm 8$  & $c_1 = 673 \pm 7$      & $c_1 = 672 \pm 8$   & $c_1 = 673 \pm 7$   \\
$O_2 = \ell^i s^i$                	                         & $c_2 = 18 \pm 19$   & $c_2 = 17 \pm 18$    & $c_2 = 19 \pm 9$    & $c_2 = 20 \pm 9$ \\
$O_3 = \frac{1}{N_c}S^iS^i$                                      & $c_3 = 121 \pm 59$  & $c_3 = 115 \pm 46$   & $c_3 = 120 \pm 58$    & $c_3 = 112 \pm 42$\\
$O_4 = \frac{1}{N_c}\left[T^aT^a-\frac{1}{12}N_c(N_c+6)\right]$  & $c_4 = 202 \pm 41$  & $c_4 = 200\pm 40$   & $c_4 = 205\pm 27$   & $c_4 = 205\pm 27$  \\
$O_5 =  \frac{3}{N_c} L^{i} T^{a} G^{ia}$                  & $c_5 = 1 \pm 13$     &   $c_5 = 2 \pm 12$  &   \\ 
$O_6 =  \frac{15}{N_c} L^{(2)ij} G^{ia} G^{ja}$                  & $c_6 = 1 \pm 6$     &   &   $c_6 = 1 \pm 5$ \\ 
\hline
$B_1 = -\mathcal{S}$                                             & $d_1 = 108 \pm 93$  & $d_1 = 108 \pm 92$ & $d_1 = 109 \pm 93$  & $d_1 = 108 \pm 92$  \\     
\hline                  
$\chi_{\mathrm{dof}}^2$                                          &  $1.23$             & $0.93$   & $0.93$      & $0.75$\\
\hline \hline
\end{tabular}}

\end{table*}
%%%%%%%%%%%%%%%%%%%%%%%%%%%%%%%%%%%%%%%%%%%%%%%%%%%%%%%%%%%

The values of the coefficients $c_i$ and $d_i$
which encode the QCD dynamics are determined from numerical fits to data.
Table  \ref{operators}  gives the list of  $O_i$  and $B_i$ operators together with  their coefficients,  which
 we believe to be the most relevant for the present study.  
The choice is based on our previous experience with the $N$ = 1 band \, \cite{Matagne:2011fr}.
In this table
%   \ref{operators} 
the first nontrivial operator is the spin-orbit operator $O_2$.
In the spirit of the Hartree picture \cite{WITTEN}
we identify the 
spin-orbit operator with the single-particle operator
\begin{equation}\label{spinorbit}
\ell \cdot s = \sum^{N_c}_{i=1} \ell(i) \cdot s(i),
\end{equation}
the matrix elements of which are of order $N^0_c$.
For simplicity  we ignore 
the two-body part of the spin-orbit operator, denoted by
$1/N_c\left(\ell \cdot S_c\right)$ in Ref. \cite{Carlson:1998vx},
as being of a lower order  (we remind that the lower case operators $\ell(i)$  act on
the excited quark and  $S_c$ is  the core spin operator).

The spin operator $O_3$ and the flavor operator $O_4$ are two-body and linearly independent. 
The expectation values of $O_3$ are simply equal to $\frac{1}{N_c} S ( S + 1 )$ where $S$ is the spin 
of the whole system. For nonstrange baryons the eigenvalue of $O_4$ is $\frac{1}{N_c} I ( I + 1 )$
where $I$ is the isospin. For the flavor singlet $\Lambda$ the eigenvalue is $-(2 N_c + 3)/4N_c $, favourably negative, 
as shown in Ref.   \cite{Matagne:2012tm}.

Note that the definition of the operator $O_4$,  indicated in Table \ref{operators},
 is such as to recover the matrix elements of
the usual $1/N_c (T^a T^a)$  in SU(4), by subtracting  $N_c (N_c+6)/12$.
This is understood  by using Eq. (30) of Ref. \cite{Matagne:2008kb} for the matrix 
elements of $1/N_c (T^a T^a)$ extended to SU(6). 
Then, it turns out that  
the expectation values  of  $O_4$   are positive for octets and decuplets 
 and of order  $N^{-1}_c$, as in SU(4), and negative   and of order $N^0_c$  for flavor singlets.

The operators  $O_5$  and $O_6$ are also two-body, which means that they carry a 
factor $1/N_c$  in the definition.  However,  as $G^{ia}$
sums coherently, it introduces an extra factor $N_c$ and makes all the matrix 
elements of $O_6$ of order $N^0_c$ \cite{Matagne:2008kb}. 
These matrix elements are obtained from the formulas (B2) and (B4) of Ref. \cite{Matagne:2011fr}
where the multiplet $[70,1^-]$ has been discussed.  Interestingly,  when $N_c$ = 3, 
the contribution of $O_5$ cancels out for flavor singlets,   like 
for $\ell$ = 1 \cite{Matagne:2011fr}. This property follows from the analytic form of the isoscalar factors 
given in Ref. \cite{Matagne:2011fr}.

We remind that 
the SU(6) generators $S^i$, $T^a$ and $G^{ia}$  and the $O(3)$ generators $L^i$ of Eq. (\ref{TENSOR}) act on the
total wave function of the $N_c$ system of quarks as proposed in Refs.  \cite{Matagne:2006dj}, 
\cite{Matagne:2008kb} and  \cite{Matagne:2011fr}.  The advantage of this procedure over the standard one, where the system is 
separated into a ground state core + an excited quark,
is that the number of relevant operators needed in the fit is usually smaller than the number of data and 
it allows a better understanding of their role in the mass formula, in  particular the role of the isospin operator $O_4$ 
which has always been omitted in the symmetric core + excited quark procedure. 
We should also mention that in our  approach the permutation symmetry is exact \cite{Matagne:2006dj}.

Among the operators containing angular momentum components, besides the spin-orbit, we have included the operators 
$O_5$ and $O_6$, to check whether or not  they bring feeble  
contributions,  as it was the case in the $N$ = 1 band.  From  Table \ref{operators}
one can see that their coefficients are indeed negligible either included together as in Fit 1 or separately as in Fit 2 and 3.
Thus in the expansion series, besides $O_1$, proportional to $N_c$, the most dominant operators are the pure spin $O_3$ and 
the pure isospin $O_4$.

\begin{figure}
\begin{center}
\input{c1.tex}
\end{center}
\caption{The coefficient $c_1$ as a function of the band number $N$: $N$ = 1 Ref. \cite{Matagne:2011fr},
$N$ = 2 Ref. \cite{Goity:2003ab} for $[{\bf 56}, 2^+]$ and Ref.  \cite{Matagne:2005gd} for $[{\bf 70}, \ell^+]$,
 $N$ = 3 Ref. \cite{Matagne:2012tm}, $N$ = 4 Ref. \cite{Matagne:2004pm}. The straight line is drawn to guide the eye.}
\end{figure}
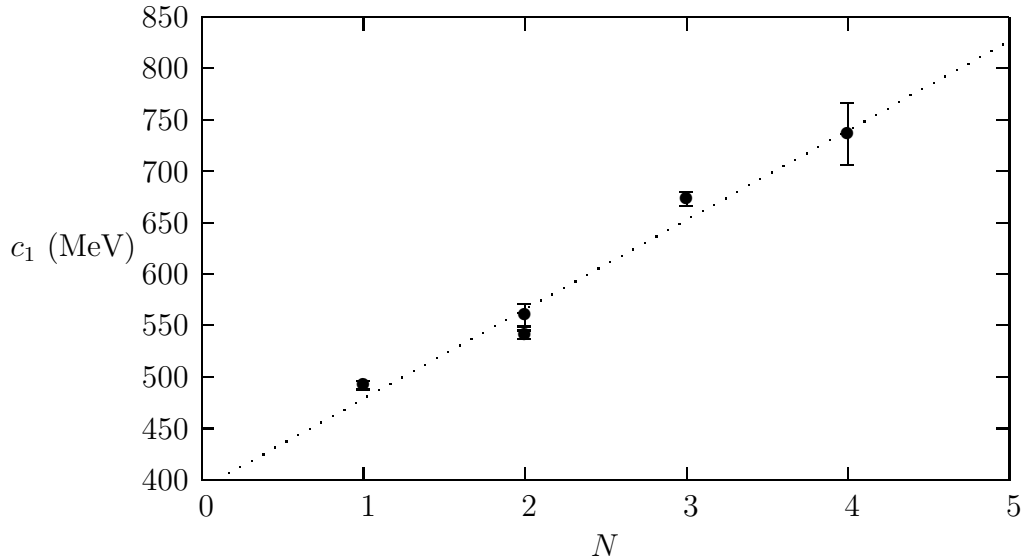
\begin{figure}
\begin{center}
\input{c2.tex}
\end{center}
\caption{Same as Figure 1 but for the coefficient $c_2$.}
\end{figure}
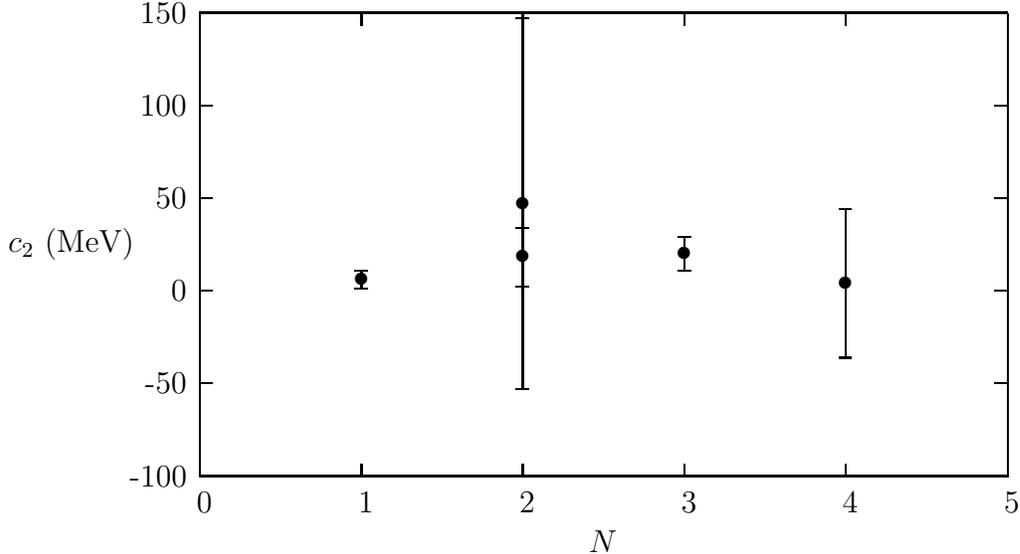

\section{Global results}
The above analysis helps us to complete previous results for $N$ = 1, 2 and 4  with the values of $c_i$ obtained 
for $N$ = 3. Therefore we can draw now a complete  picture of the dependence of the coefficients $c_1$ 
and $c_2$ on $N$ in analogy to Ref. \cite{Matagne:2005gd} where results for $N$ = 3 were missing. 
The new pictures are shown in Figs. 1 and 2. One can see that the values of  $c_1$ follow nearly a straight line which
can give rise to a Regge trajectory. Remember that $c_1$ describes the bulk content of the baryon mass, $c_1 N_c$ being the most dominant 
mass term. In a quark model language it represents the kinetic plus the confinement energy. As
as discussed in Refs. \cite{Semay:2007cv,Buisseret:2008tq} the band number $N$ also emerges
from the spin independent part of a semi-relativistic quark model. If this part contributes to
the total mass by a quantity denoted by $M_0$, then one can make the identification
\begin{equation}
c^2_1 = M^2_0/9
\end{equation}
when $N_c$ = 3. In this way one can compare the Regge trajectory obtainable from the above results with
that of a standard constituent quark model. It turns out that they are close to each other \cite{Semay:2007cv,Buisseret:2008tq}.
and the value obtained here for $c_1$ at  $N$ = 3, missing in the previous work, is entirely compatible with
the previous picture.    

The behaviour of $c_2$  shows that the spin-orbit operator contributes very little to the mass, at 
all energies, in agreement to quark models, where it is usually neglected. Note that the behaviour of 
$c_2$ in Fig. 2 is slightly different from that of \cite{Matagne:2005gd}, because we presently take
the value of $c_2$ at $N$ = 1 from  Ref. \cite{Matagne:2011fr} (Fit 3 giving the lowest $\chi_{\mathrm{dof}}^2$) 
for consistency with our 
treatment, instead of that of Ref. \cite{Carlson:1998vx}, based on the ground state core + excited quark, 
the only available at the time the paper \cite{Matagne:2005gd} was published. 

We refrain ourselves from presenting the global picture of 
$c_3$, the spin term coefficient, because the results for positive parity mixed symmetric states are obtained on the one hand in the 
core + excited quark approach, where the isospin term is missing and on the other hand,
for negative parity states where it is present, our approach is used. This term competes with the spin term. We plan to 
reanalyze the $[{\bf 70}, \ell^+]$ multiplets before drawing a complete picture of $c_3$.

%%%%%%%%%%%%%%%%%%%%%%%%%%%%%%%%%%%%%%%%%%%%%%%%%%%%%%%%%%%%%%%%%%%%%

\section{Conclusions}\label{se:concl}
We have used a procedure which allows to write the mass formula by 
using a small number of linearly independent operators for spin-flavour  mixed symmetric states 
of SU(6).  The numerical fits of the dynamical coefficients in the mass formula 
for $N$ = 3 band resonances show that the pure spin and pure flavor terms are dominant
in the $1/N_c$ expansion,
like for $N$ = 1 resonances. This proves that the isospin term cannot be neglected,
as it was the case in the ground state + excited quark procedure. We have shown 
the dependence of the  dynamical coefficients $c_1$ and $c_2$ as a function of the band number $N$ or 
alternatively of the excitation energy for $N$ = 1, 2, 3 and 4 bands.

%%%%%%%%%%%%%%%%%%%%%%%%%%%%%%%%%%%%%%%%%%%%%%%%%%%%%%%%%%%%%%%%%%%%%%%%%%%%%%%%
 
\end{document}

%% file: c1.tex
% GNUPLOT: LaTeX picture
\setlength{\unitlength}{0.240900pt}
\ifx\plotpoint\undefined\newsavebox{\plotpoint}\fi
\sbox{\plotpoint}{\rule[-0.200pt]{0.400pt}{0.400pt}}%
\begin{picture}(1500,900)(0,0)
\sbox{\plotpoint}{\rule[-0.200pt]{0.400pt}{0.400pt}}%
\put(171.0,131.0){\rule[-0.200pt]{4.818pt}{0.400pt}}
\put(151,131){\makebox(0,0)[r]{ 400}}
\put(1419.0,131.0){\rule[-0.200pt]{4.818pt}{0.400pt}}
\put(171.0,212.0){\rule[-0.200pt]{4.818pt}{0.400pt}}
\put(151,212){\makebox(0,0)[r]{ 450}}
\put(1419.0,212.0){\rule[-0.200pt]{4.818pt}{0.400pt}}
\put(171.0,293.0){\rule[-0.200pt]{4.818pt}{0.400pt}}
\put(151,293){\makebox(0,0)[r]{ 500}}
\put(1419.0,293.0){\rule[-0.200pt]{4.818pt}{0.400pt}}
\put(171.0,374.0){\rule[-0.200pt]{4.818pt}{0.400pt}}
\put(151,374){\makebox(0,0)[r]{ 550}}
\put(1419.0,374.0){\rule[-0.200pt]{4.818pt}{0.400pt}}
\put(171.0,455.0){\rule[-0.200pt]{4.818pt}{0.400pt}}
\put(151,455){\makebox(0,0)[r]{ 600}}
\put(1419.0,455.0){\rule[-0.200pt]{4.818pt}{0.400pt}}
\put(171.0,535.0){\rule[-0.200pt]{4.818pt}{0.400pt}}
\put(151,535){\makebox(0,0)[r]{ 650}}
\put(1419.0,535.0){\rule[-0.200pt]{4.818pt}{0.400pt}}
\put(171.0,616.0){\rule[-0.200pt]{4.818pt}{0.400pt}}
\put(151,616){\makebox(0,0)[r]{ 700}}
\put(1419.0,616.0){\rule[-0.200pt]{4.818pt}{0.400pt}}
\put(171.0,697.0){\rule[-0.200pt]{4.818pt}{0.400pt}}
\put(151,697){\makebox(0,0)[r]{ 750}}
\put(1419.0,697.0){\rule[-0.200pt]{4.818pt}{0.400pt}}
\put(171.0,778.0){\rule[-0.200pt]{4.818pt}{0.400pt}}
\put(151,778){\makebox(0,0)[r]{ 800}}
\put(1419.0,778.0){\rule[-0.200pt]{4.818pt}{0.400pt}}
\put(171.0,859.0){\rule[-0.200pt]{4.818pt}{0.400pt}}
\put(151,859){\makebox(0,0)[r]{ 850}}
\put(1419.0,859.0){\rule[-0.200pt]{4.818pt}{0.400pt}}
\put(171.0,131.0){\rule[-0.200pt]{0.400pt}{4.818pt}}
\put(171,90){\makebox(0,0){ 0}}
\put(171.0,839.0){\rule[-0.200pt]{0.400pt}{4.818pt}}
\put(425.0,131.0){\rule[-0.200pt]{0.400pt}{4.818pt}}
\put(425,90){\makebox(0,0){ 1}}
\put(425.0,839.0){\rule[-0.200pt]{0.400pt}{4.818pt}}
\put(678.0,131.0){\rule[-0.200pt]{0.400pt}{4.818pt}}
\put(678,90){\makebox(0,0){ 2}}
\put(678.0,839.0){\rule[-0.200pt]{0.400pt}{4.818pt}}
\put(932.0,131.0){\rule[-0.200pt]{0.400pt}{4.818pt}}
\put(932,90){\makebox(0,0){ 3}}
\put(932.0,839.0){\rule[-0.200pt]{0.400pt}{4.818pt}}
\put(1185.0,131.0){\rule[-0.200pt]{0.400pt}{4.818pt}}
\put(1185,90){\makebox(0,0){ 4}}
\put(1185.0,839.0){\rule[-0.200pt]{0.400pt}{4.818pt}}
\put(1439.0,131.0){\rule[-0.200pt]{0.400pt}{4.818pt}}
\put(1439,90){\makebox(0,0){ 5}}
\put(1439.0,839.0){\rule[-0.200pt]{0.400pt}{4.818pt}}
\put(171.0,131.0){\rule[-0.200pt]{0.400pt}{175.375pt}}
\put(171.0,131.0){\rule[-0.200pt]{305.461pt}{0.400pt}}
\put(1439.0,131.0){\rule[-0.200pt]{0.400pt}{175.375pt}}
\put(171.0,859.0){\rule[-0.200pt]{305.461pt}{0.400pt}}
\put(-30,495){\makebox(0,0){$c_1$ (MeV)}}
\put(805,29){\makebox(0,0){$N$}}
\put(425.0,273.0){\rule[-0.200pt]{0.400pt}{3.132pt}}
\put(415.0,273.0){\rule[-0.200pt]{4.818pt}{0.400pt}}
\put(415.0,286.0){\rule[-0.200pt]{4.818pt}{0.400pt}}
\put(678.0,353.0){\rule[-0.200pt]{0.400pt}{3.132pt}}
\put(668.0,353.0){\rule[-0.200pt]{4.818pt}{0.400pt}}
\put(668.0,366.0){\rule[-0.200pt]{4.818pt}{0.400pt}}
\put(678.0,372.0){\rule[-0.200pt]{0.400pt}{8.672pt}}
\put(668.0,372.0){\rule[-0.200pt]{4.818pt}{0.400pt}}
\put(668.0,408.0){\rule[-0.200pt]{4.818pt}{0.400pt}}
\put(932.0,561.0){\rule[-0.200pt]{0.400pt}{5.541pt}}
\put(922.0,561.0){\rule[-0.200pt]{4.818pt}{0.400pt}}
\put(922.0,584.0){\rule[-0.200pt]{4.818pt}{0.400pt}}
\put(1185.0,626.0){\rule[-0.200pt]{0.400pt}{23.367pt}}
\put(1175.0,626.0){\rule[-0.200pt]{4.818pt}{0.400pt}}
\put(425,280){\makebox(0,0){$\bullet$}}
\put(678,359){\makebox(0,0){$\bullet$}}
\put(678,390){\makebox(0,0){$\bullet$}}
\put(932,573){\makebox(0,0){$\bullet$}}
\put(1185,675){\makebox(0,0){$\bullet$}}
\put(1175.0,723.0){\rule[-0.200pt]{4.818pt}{0.400pt}}
\put(195.00,131.00){\usebox{\plotpoint}}
\put(213.07,141.19){\usebox{\plotpoint}}
\put(231.35,151.03){\usebox{\plotpoint}}
\put(249.62,160.87){\usebox{\plotpoint}}
\put(267.77,170.95){\usebox{\plotpoint}}
\put(285.52,181.70){\usebox{\plotpoint}}
\put(303.78,191.57){\usebox{\plotpoint}}
\put(322.05,201.41){\usebox{\plotpoint}}
\put(340.28,211.33){\usebox{\plotpoint}}
\put(358.37,221.51){\usebox{\plotpoint}}
\put(376.64,231.35){\usebox{\plotpoint}}
\put(394.92,241.19){\usebox{\plotpoint}}
\put(412.58,252.05){\usebox{\plotpoint}}
\put(430.77,262.03){\usebox{\plotpoint}}
\put(449.04,271.87){\usebox{\plotpoint}}
\put(467.29,281.75){\usebox{\plotpoint}}
\put(485.36,291.96){\usebox{\plotpoint}}
\put(503.63,301.80){\usebox{\plotpoint}}
\put(521.75,311.92){\usebox{\plotpoint}}
\put(539.56,322.58){\usebox{\plotpoint}}
\put(557.79,332.50){\usebox{\plotpoint}}
\put(576.06,342.34){\usebox{\plotpoint}}
\put(594.33,352.19){\usebox{\plotpoint}}
\put(612.38,362.43){\usebox{\plotpoint}}
\put(630.65,372.27){\usebox{\plotpoint}}
\put(648.80,382.34){\usebox{\plotpoint}}
\put(666.76,392.72){\usebox{\plotpoint}}
\put(684.80,402.97){\usebox{\plotpoint}}
\put(703.08,412.81){\usebox{\plotpoint}}
\put(721.35,422.65){\usebox{\plotpoint}}
\put(739.54,432.65){\usebox{\plotpoint}}
\put(757.67,442.75){\usebox{\plotpoint}}
\put(775.51,453.35){\usebox{\plotpoint}}
\put(793.78,463.19){\usebox{\plotpoint}}
\put(811.82,473.44){\usebox{\plotpoint}}
\put(830.10,483.28){\usebox{\plotpoint}}
\put(848.37,493.12){\usebox{\plotpoint}}
\put(866.58,503.09){\usebox{\plotpoint}}
\put(884.69,513.22){\usebox{\plotpoint}}
\put(902.52,523.82){\usebox{\plotpoint}}
\put(920.80,533.66){\usebox{\plotpoint}}
\put(938.85,543.91){\usebox{\plotpoint}}
\put(957.12,553.76){\usebox{\plotpoint}}
\put(975.39,563.60){\usebox{\plotpoint}}
\put(993.67,573.44){\usebox{\plotpoint}}
\put(1011.50,584.00){\usebox{\plotpoint}}
\put(1029.52,594.28){\usebox{\plotpoint}}
\put(1047.79,604.12){\usebox{\plotpoint}}
\put(1066.07,613.96){\usebox{\plotpoint}}
\put(1084.11,624.21){\usebox{\plotpoint}}
\put(1102.38,634.05){\usebox{\plotpoint}}
\put(1120.66,643.89){\usebox{\plotpoint}}
\put(1138.55,654.37){\usebox{\plotpoint}}
\put(1156.51,664.74){\usebox{\plotpoint}}
\put(1174.78,674.58){\usebox{\plotpoint}}
\put(1193.06,684.42){\usebox{\plotpoint}}
\put(1211.10,694.67){\usebox{\plotpoint}}
\put(1229.38,704.51){\usebox{\plotpoint}}
\put(1247.63,714.39){\usebox{\plotpoint}}
\put(1265.38,725.14){\usebox{\plotpoint}}
\put(1283.53,735.21){\usebox{\plotpoint}}
\put(1301.80,745.05){\usebox{\plotpoint}}
\put(1320.08,754.89){\usebox{\plotpoint}}
\put(1338.33,764.77){\usebox{\plotpoint}}
\put(1356.39,774.98){\usebox{\plotpoint}}
\put(1374.67,784.82){\usebox{\plotpoint}}
\put(1392.50,795.43){\usebox{\plotpoint}}
\put(1410.59,805.60){\usebox{\plotpoint}}
\put(1428.82,815.52){\usebox{\plotpoint}}
\put(1439,821){\usebox{\plotpoint}}
\put(171.0,131.0){\rule[-0.200pt]{0.400pt}{175.375pt}}
\put(171.0,131.0){\rule[-0.200pt]{305.461pt}{0.400pt}}
\put(1439.0,131.0){\rule[-0.200pt]{0.400pt}{175.375pt}}
\put(171.0,859.0){\rule[-0.200pt]{305.461pt}{0.400pt}}
\end{picture}

%% file: c2.tex
% GNUPLOT: LaTeX picture
\setlength{\unitlength}{0.240900pt}
\ifx\plotpoint\undefined\newsavebox{\plotpoint}\fi
\sbox{\plotpoint}{\rule[-0.200pt]{0.400pt}{0.400pt}}%
\begin{picture}(1500,900)(0,0)
\sbox{\plotpoint}{\rule[-0.200pt]{0.400pt}{0.400pt}}%
\put(171.0,131.0){\rule[-0.200pt]{4.818pt}{0.400pt}}
\put(151,131){\makebox(0,0)[r]{-100}}
\put(1419.0,131.0){\rule[-0.200pt]{4.818pt}{0.400pt}}
\put(171.0,277.0){\rule[-0.200pt]{4.818pt}{0.400pt}}
\put(151,277){\makebox(0,0)[r]{-50}}
\put(1419.0,277.0){\rule[-0.200pt]{4.818pt}{0.400pt}}
\put(171.0,422.0){\rule[-0.200pt]{4.818pt}{0.400pt}}
\put(151,422){\makebox(0,0)[r]{ 0}}
\put(1419.0,422.0){\rule[-0.200pt]{4.818pt}{0.400pt}}
\put(171.0,568.0){\rule[-0.200pt]{4.818pt}{0.400pt}}
\put(151,568){\makebox(0,0)[r]{ 50}}
\put(1419.0,568.0){\rule[-0.200pt]{4.818pt}{0.400pt}}
\put(171.0,713.0){\rule[-0.200pt]{4.818pt}{0.400pt}}
\put(151,713){\makebox(0,0)[r]{ 100}}
\put(1419.0,713.0){\rule[-0.200pt]{4.818pt}{0.400pt}}
\put(171.0,859.0){\rule[-0.200pt]{4.818pt}{0.400pt}}
\put(151,859){\makebox(0,0)[r]{ 150}}
\put(1419.0,859.0){\rule[-0.200pt]{4.818pt}{0.400pt}}
\put(171.0,131.0){\rule[-0.200pt]{0.400pt}{4.818pt}}
\put(171,90){\makebox(0,0){ 0}}
\put(171.0,839.0){\rule[-0.200pt]{0.400pt}{4.818pt}}
\put(425.0,131.0){\rule[-0.200pt]{0.400pt}{4.818pt}}
\put(425,90){\makebox(0,0){ 1}}
\put(425.0,839.0){\rule[-0.200pt]{0.400pt}{4.818pt}}
\put(678.0,131.0){\rule[-0.200pt]{0.400pt}{4.818pt}}
\put(678,90){\makebox(0,0){ 2}}
\put(678.0,839.0){\rule[-0.200pt]{0.400pt}{4.818pt}}
\put(932.0,131.0){\rule[-0.200pt]{0.400pt}{4.818pt}}
\put(932,90){\makebox(0,0){ 3}}
\put(932.0,839.0){\rule[-0.200pt]{0.400pt}{4.818pt}}
\put(1185.0,131.0){\rule[-0.200pt]{0.400pt}{4.818pt}}
\put(1185,90){\makebox(0,0){ 4}}
\put(1185.0,839.0){\rule[-0.200pt]{0.400pt}{4.818pt}}
\put(1439.0,131.0){\rule[-0.200pt]{0.400pt}{4.818pt}}
\put(1439,90){\makebox(0,0){ 5}}
\put(1439.0,839.0){\rule[-0.200pt]{0.400pt}{4.818pt}}
\put(171.0,131.0){\rule[-0.200pt]{0.400pt}{175.375pt}}
\put(171.0,131.0){\rule[-0.200pt]{305.461pt}{0.400pt}}
\put(1439.0,131.0){\rule[-0.200pt]{0.400pt}{175.375pt}}
\put(171.0,859.0){\rule[-0.200pt]{305.461pt}{0.400pt}}
\put(-30,495){\makebox(0,0){$c_2$ (MeV)}}
\put(805,29){\makebox(0,0){$N$}}
\put(425.0,425.0){\rule[-0.200pt]{0.400pt}{6.986pt}}
\put(415.0,425.0){\rule[-0.200pt]{4.818pt}{0.400pt}}
\put(415.0,454.0){\rule[-0.200pt]{4.818pt}{0.400pt}}
\put(678.0,428.0){\rule[-0.200pt]{0.400pt}{22.404pt}}
\put(668.0,428.0){\rule[-0.200pt]{4.818pt}{0.400pt}}
\put(668.0,521.0){\rule[-0.200pt]{4.818pt}{0.400pt}}
\put(678.0,268.0){\rule[-0.200pt]{0.400pt}{140.204pt}}
\put(668.0,268.0){\rule[-0.200pt]{4.818pt}{0.400pt}}
\put(668.0,850.0){\rule[-0.200pt]{4.818pt}{0.400pt}}
\put(932.0,454.0){\rule[-0.200pt]{0.400pt}{12.768pt}}
\put(922.0,454.0){\rule[-0.200pt]{4.818pt}{0.400pt}}
\put(922.0,507.0){\rule[-0.200pt]{4.818pt}{0.400pt}}
\put(1185.0,317.0){\rule[-0.200pt]{0.400pt}{56.130pt}}
\put(1175.0,317.0){\rule[-0.200pt]{4.818pt}{0.400pt}}
\put(425,440){\makebox(0,0){$\bullet$}}
\put(678,475){\makebox(0,0){$\bullet$}}
\put(678,559){\makebox(0,0){$\bullet$}}
\put(932,480){\makebox(0,0){$\bullet$}}
\put(1185,434){\makebox(0,0){$\bullet$}}
\put(1175.0,550.0){\rule[-0.200pt]{4.818pt}{0.400pt}}
\put(171.0,131.0){\rule[-0.200pt]{0.400pt}{175.375pt}}
\put(171.0,131.0){\rule[-0.200pt]{305.461pt}{0.400pt}}
\put(1439.0,131.0){\rule[-0.200pt]{0.400pt}{175.375pt}}
\put(171.0,859.0){\rule[-0.200pt]{305.461pt}{0.400pt}}
\end{picture}